\documentclass[12pt]{article}
\usepackage{amsbsy,epsf}
\def\noi{\noindent}
\def\beq{\begin{equation}}
\def\eeq{\end{equation}}
\newcommand{\bea}{\begin{eqnarray}}
\newcommand{\eea}{\end{eqnarray}}
\newcommand{\nn}{\nonumber}
\font\boldgreek=cmmib10
\mathchardef\mysigma="091B
\font \sixteenrm = cmr10 at 16 pt
\def\lsim{\raise0.3ex\hbox{$<$\kern-0.75em\raise-1.1ex\hbox{$\sim$}}}
\def\gsim{\raise0.3ex\hbox{$>$\kern-0.75em\raise-1.1ex\hbox{$\sim$}}}
\begin{document}
\begin{center}
{\large \bf Decay constants in the heavy quark limit} \\
{\large \bf in models \`a la Bakamjian and Thomas} \\

\vskip 5 truemm

{\bf V. Mor\'enas}\footnote{E-mail : morenas@clermont.in2p3.fr}\\

Laboratoire de Physique Corpusculaire \\
Université Blaise Pascal - CNRS/IN2P3 F-63000 Aubi{\`e}re Cedex, France \\

\vskip 3 truemm
{\bf A. Le Yaouanc\footnote{E-mail : leyaouan@qcd.th.u-psud.fr}, L.
Oliver\footnote{E-mail : oliver@qcd.th.u-psud.fr}, O. P\`ene\footnote{E-mail :
pene@qcd.th.u-psud.fr}
and J.-C. Raynal}\\

Laboratoire de Physique Th\'eorique et Hautes \'Energies\footnote{Laboratoire
associ\'e au Centre National de la Recherche Scientifique - URA D0063}\\
Universit\'e de Paris XI, B\^atiment 210,
F-91405 Orsay Cedex, France
\\
\end{center}
\vskip 1 truecm

\begin{abstract}
In quark models \`a la Bakamjian and Thomas, that yield covariance and Isgur-Wise
scaling of form
factors in the heavy quark limit, we compute the decay constants $f^{(n)}$ and
$f^{(n)}_{1/2}$ of
$S$-wave and $P$-wave mesons composed of heavy and light quarks. Heavy quark
limit scaling
$\sqrt{M}\  f =$ Cst. is obtained, and it is shown that this class of models
satisfies the sum rules
involving decay constants and Isgur-Wise functions recently formulated by us in
the heavy quark
limit of QCD. Moreover, the model also satisfies the selection rules of the
type $f^{(n)}_{3/2} = 0$
that must hold in this limit. We discuss different Ans\"atze for the dynamics of
the mass operator at
rest. For non-relativistic kinetic energies $\displaystyle{{p^2 \over 2m}}$
the decay constants are
finite even if the potential $V(r)$ has a Coulomb part. For the relativistic
form $\sqrt{p^2 +
m^2}$, the $S$-wave decay constants diverge if there is a Coulomb singularity.
Using
phenomenological models of the spectrum with relativistic kinetic energy and
regularized short
distance part (Godfrey-Isgur model or Richardson potential of Colangelo et
al.), that yield $\rho^2
\cong 1$ for the elastic Isgur-Wise function, we compute the decay constants in
the heavy quark
limit, and obtain $f_B \cong$ 300 MeV, of the same order although slightly
smaller than in the
static limit of lattice QCD. We find the decay constants of $D^{**}$ with $j =
\displaystyle{{1 \over
2}}$  of the same order of magnitude. The convergence of the heavy quark limit
sum rules is also
studied. \end{abstract}
\vspace{1 truecm}
\noi LPTHE 97-23 \par
\noi  PCCF RI 9708 \par
\noi hep-ph/9710298 \par
\noi July 1997 \par
\newpage
\pagestyle{plain}
\baselineskip=18pt
\section{Introduction}
\hspace*{\parindent}
In a recent paper \cite{1r} we have shown that quark models of hadrons with a
fixed number of
constituents, based on the Bakamjian-Thomas (BT) formalism \cite{2r}, yield
form factors that are
covariant and satisfy Isgur-Wise (IW) scaling \cite{3r}  in the heavy mass
limit for one of the
quarks. In this class of models a lower bound is predicted for the slope of the
heavy meson elastic
IW function $\rho^2 > \displaystyle{{3 \over 4}}$ . Moreover, the model
satisfies the
Bjorken-Isgur-Wise sum rule \cite{4r} that relates the slope of the IW function
to the $P$-wave form
factors $\tau_{1/2}(w)$, $\tau_{3/2}(w)$ at zero recoil \cite{5r}. We have also
explicitly computed
the $P$-wave meson wave functions and the corresponding inelastic IW functions
\cite{6r} within the
model, we have made a numerical study of $\rho^2$ in a wide class of models of
the meson spectrum
(each of them characterized by an Ansatz for the mass operator $M$, i.e. the
dynamics of the system
at rest) \cite{7r}, and a phenomenological study \cite{8r} of the elastic and
inelastic IW functions
and the corresponding $\displaystyle{{d\Gamma \over dw}}$ for $B \to D$, $D^*$,
$D^{**} \ell \nu$.
\par

The purpose of the present paper is to study the decay constants of heavy
mesons within the same
approach.

\section{Decay constants in the B-T scheme : heavy quark scaling}
\hspace*{\parindent}

We want to compute in the model the decay constants of mesons $Q\bar{q}$, where
$Q$ and $q$ are a
heavy and a light quark, defined by

\bea
&&< P(0^-)(v)|A^{qQ}_{\mu}|0 > = N \ f_P \ \sqrt{M} \ v_{\mu} \nn \\
&&< V(1^-)(v,\varepsilon )|V^{qQ}_{\mu}|0 > = N \ f_V \ \sqrt{M} \
\varepsilon^*_{\mu}
\label{1e}
\eea

\noi for $S$ states, where $N = \displaystyle{{1 \over\sqrt{2v^0}}}$  and $M$
is the mass of the
corresponding bound state ($M = M_P$ or $M = M_V$), and similarly :

\bea
&&< ^{1/2}0^+(v)|V^{qQ}_{\mu}|0 > = N \ f^{(1/2)} \sqrt{M} \ v_{\mu} \nn \\
&&< ^{1/2}1^+(v,\varepsilon )|A^{qQ}_{\mu}|0 > = N \ g^{(1/2)} \sqrt{M} \
\varepsilon^*_{\mu} \nn
\\
&&< ^{3/2}1^+(v,\varepsilon )|A^{qQ}_{\mu}|0 > = N \ g^{(3/2)} \sqrt{M} \
\varepsilon^*_{\mu}
\label{2e}
\eea

\noi for $P$ states. The model satisfies the heavy quark limit relations (we
use the phase convention
for states of ref. \cite{6r})

\beq
			f_P = f_V = f
\label{3e}
\eeq

\beq
		g^{(1/2)} =  - f^{(1/2)} 	\qquad	g^{(3/2)} = 0 \ \ \ .
\label{4e}
\eeq

\noi In terms of the internal wave function we get (the superindex $(n)$ means
any radial
excitation), for $S$-wave mesons~:

\beq
\sqrt{M} f^{(n)} = \sqrt{N_c} \sqrt{2}  \int {d{\bf p}_2 \over (2\pi)^3} {1
\over p^0_2}
\sqrt{(p_2 \cdot v)(p_2 \cdot v+m^2)} \ \varphi^{(n)}({\bf k}_2)
\label{5e}
\eeq

\noi where the wave functions at rest of $0^-$ mesons are given by

\bea
&&\varphi_{s_1,s_2}^{(n)}({\bf k}_2) = {i \over \sqrt{2}}  (\sigma^2)_{s_1,s_2}
\varphi^{(n)}({\bf k}_2)	\nn \\
&&\int {d{\bf p} \over (2\pi)^3} |\varphi^{(n)}({\bf p})|^2 = 1
\label{6e}
\eea

\noi and for $P$ wave mesons :

\beq
\sqrt{M} f^{(n)}_{1/2} = \sqrt{{2 \over 3}} \sqrt{N_c} \int {d{\bf p}_2 \over
(2\pi)^3}
{1 \over p^0_2} {\sqrt{p_2 \cdot v} \over \sqrt{p_2 \cdot v + m_2}}
\varphi^{(n)}_{1/2}({\bf k}_2) \left [ (p_2 \cdot v)^2 - m_2^2 \right ]
\label{7e}
\eeq

\noi where the wave function at rest of $0^+$ mesons is given by \cite{5r}

\bea
\varphi_{s_1,s_2}^{(n)}({\bf k}_2) &=& - {i \over \sqrt{6}} \left [ \left (
\boldsymbol{\sigma} \cdot
{\bf k}_2 \right ) \sigma^2 \right ]_{s_1,s_2} \varphi^{(n)}_{1/2}({\bf k}_2)
\nn \\
&&\int {d{\bf p} \over (2\pi)^3}  {p^2 \over 3} |\varphi^{(n)}_{1/2}({\bf
p})|^2 = 1 \ \ \ .
\label{8e}
\eea

\noi In these expressions,

\beq
			p_2 = B_v\ k_2
\label{9e}
\eeq

\noi where $B_v$ is the boost operator. The expressions (\ref{5e}) and
(\ref{7e}) are Lorentz
invariant since, due to the \underbar{rotational invariance} of
$\varphi^{(n)}({\bf k}_2)$,
$\varphi^{(n)}_{1/2}({\bf k}_2)$, they are functions of $v^2=1$. One can take
simply the rest frame
$v = (1,{\bf 0})$. \par

	Let us give a brief outline of the calculation. Within the model \cite{1r},
the internal wave
function in motion of a $S$-wave meson  with center-of-mass momentum ${\bf P}$
($n$ denotes the
radial excitation) is given in terms of~:

\beq
\Psi^{(n)}_{s_1s_2}({\bf P} - {\bf p}_2,{\bf p}_2) = \sqrt{{\Sigma p^0_j \over
M_0}}  \prod_{i=1,n}
{\sqrt{k^0_i} \over \sqrt{p^0_i}} \sum_{\{s'_i\}} \prod_{i=1,2} \left [
D_i({\bf R}_i)
\right ]_{s_i s'_i} \varphi^{(n)}_{s'_1s'_2}({\bf k}_2)
\label{10e}
\eeq

\noi where $\varphi^{(n)}_{s_1s_2}({\bf k}_2)$ is the internal wave function at
rest (\ref{5e}), and
${\bf R}_i$ are the Wigner rotations~:

\beq
	{\bf R}_i = {\bf B}^{-1}_{p_i} \ {\bf B}_{\Sigma p_j} \  {\bf B}_{k_i}		\ \ \
{}.
\label{11e}
\eeq

\noi The $q\bar{Q}$ to vacuum matrix element of a current, e.g. $J =
\gamma_{\mu} \gamma_5$, reads

\beq
O_{s_1s_2}({\bf P} - {\bf p}_2,{\bf p}_2) = < 0|J|{\bf P} - {\bf p}_2 s_1 ;
{\bf p}_2 s_2 > =
\left [u_{s_1}({\bf P} - {\bf p}_2) \right ]^t i\gamma^2 \gamma^0 \gamma_{\mu}
\gamma_5
u_{s_2}({\bf p}_2)
\label{12e}
\eeq

The matrix element of interest to us

\beq
< 0|A^{\mu}|P^{(n)} > = \sqrt{N_c} \int {d{\bf p}_2 \over (2\pi)^3}
\sum_{\{s_i\}}
O_{s_1s_2}({\bf P} - {\bf p}_2,{\bf p}_2) \Psi^{(n)}_{s_2s_1}({\bf P} - {\bf
p}_2,{\bf p}_2)
\label{13e}
\eeq

\noi can be written, after some algebra, using (\ref{10e})-(\ref{12e})~:

\bea
&&< 0|A^{\mu}|P^{(n)}> = \sqrt{N_c} \int {d{\bf p}_2 \over (2\pi)^3}
\sqrt{{\Sigma p^0_j \over M_0}}
\prod_{i=1,2} {\sqrt{k^0_i} \over \sqrt{p^0_i}} \sqrt{{m_1 \over p^0_1}}
\sqrt{{m_2 \over
p^0_2}}  \varphi^{(n)}({\bf k}_2) \nn \\
&&{1 \over\sqrt{2}}  \hbox{\sixteenrm Tr} \left \{ {1 + \gamma^0 \over 2} {\bf
B}^{-1}_{p_1}
\gamma^{\mu}  {\bf B}_{p_2} {1 + \gamma^0 \over 2} {\bf B}^{-1}_{p_2} {\bf
B}_{\Sigma p_j}
{\bf B}_{k_2} {\bf B}^{-1}_{k_1} {\bf B}^{-1}_{\Sigma p_j} {\bf B}_{p_1} \right
\}	 \label{14e}
\eea

\noi and since the projector $\displaystyle{{1 + \gamma^0 \over 2}}$  commutes
with the Wigner
rotations (\ref{11e}), one can rewrite~:

\bea
&&< 0|A^{\mu}|P^{(n)}> = \sqrt{N_c} \int {d{\bf p}_2 \over (2\pi)^3}
\sqrt{{\Sigma p^0_j \over M_0}}
\prod_{i=1,2} {\sqrt{k^0_i} \over \sqrt{p^0_i}} \sqrt{{m_1 \over p^0_1}}
\sqrt{{m_2 \over
p^0_2}}  \varphi^{(n)}({\bf k}_2) \nn \\
&&{1 \over\sqrt{2}}  \hbox{\sixteenrm Tr}\left \{ \gamma^{\mu} {\bf B}_u {\bf
B}_{k_2}
{1+\gamma^0 \over 2} {\bf B}^{-1}_u {\bf B}_u {1+\gamma^0 \over 2} {\bf
B}^{-1}_{k_1}
{\bf B}^{-1}_u \right \}
\label{15e}
\eea

\noi where $u = \displaystyle{{p_1+p_2 \over \sqrt{(p_1+p_2)^2}}}$. Using the
identities~:

\bea
&&{\bf	B}_u \ {\bf B}_{k_2} {1+\gamma^0 \over 2} {\bf B}^{-1}_u = {m_2+ {/
\hskip - 2 truemm
p}_ 2 \over \sqrt{2m_2 \left ( k_2^0+m_2 \right )}} {1+{/ \hskip - 2 truemm u}
\over 2} \nn \\
&&{\bf B}_u {1+\gamma^0 \over 2} {\bf B}^{-1}_{k_1} {\bf B}^{-1}_u = {1+ {/
\hskip - 2 truemm u}
\over 2} {m_1+{/ \hskip - 2 truemm p}_1 \over \sqrt{2m_1 \left ( k_1^0 + m_1
\right ) }}
\label{16e}
\eea

\noi one obtains

\bea
&&< 0|A^{\mu}|P^{(n)} > = {\sqrt{N_c} \over 8} \int {d{\bf p}_2 \over (2\pi)^3}
 {1 \over p^0_2}
F({\bf p}_2,{\bf P}) \nn \\
&&Tr \left [ \gamma^{\mu} \left (m_2+{/ \hskip - 2 truemm p}_2 \right ) ( 1 +
{/
\hskip - 2 truemm u}) \left ( m_1+ {/ \hskip - 2 truemm p}_1 \right ) \right ]
\varphi^{(n)}({\bf k}_2)				 \label{17e} \eea

\noi with ${\bf k}^2 = {\bf B}_{u}^{-1}p^2$ and

\beq
F({\bf p}_2,{\bf P}) = \sqrt{2} {\sqrt{u^0} \over p^0_1} {\sqrt{k^0_1} \over
\sqrt{k^0_1+m_1}}
{\sqrt{k^0_2} \over \sqrt{k^0_2 +m_2}} \ \ \ .
\label{18e}
\eeq

\noi Finally, in the heavy mass limit \cite{1r}~:

\begin{eqnarray}
&u \to v		&\qquad \qquad {p_1 \over m_1}  \to v \nn \\
&\displaystyle{{k_1^0 \over m_1}}  \to 1	      &\qquad \qquad k_2^0 \to \left (
{\bf B}^{-1}_v p_2
\right )^0 = p_2 \cdot v				 \label{19e}
\end{eqnarray}

\noi one obtains :

\bea
&&< 0|A^{\mu}|P^{(n)} > = {1 \over \sqrt{2v^0}} {\sqrt{N_c} \over 2\sqrt{2}}
\int {d{\bf
p}_2 \over (2\pi)^3}  {1 \over p^0_2} {\sqrt{p_2 \cdot v} \over \sqrt{p_2 \cdot
v+m_2}} \nn \\
&&Tr \left [ \gamma^{\mu} \left ( m_2+ {/ \hskip - 2 truemm p}_2 \right ) (1+{/
\hskip - 2
truemm v}) \right ] \varphi^{(n)}({\bf k}_2)						 \label{20e}
\eea

\noi where $k_2 = {\bf B}_v^{-1}p_2$ and $\varphi^{(n)}({\bf k}_2)$ depends on
${\bf k}_2^2 =
(p_2 \cdot v)^2-m_2^2$  because of rotational invariance.  \par

Thus, the integrand (\ref{20e}) must be
proportional to $v^{\mu}$, yielding equation (\ref{5e}) for the decay constant.
This expression
exhibits the expected heavy quark limit scaling. Moreover, varying the current,
one can easily show
relations (\ref{3e}). It is also worth noticing that expression (\ref{5e})
becomes the well-known
non-relativistic expression \cite{9r} if we assume that the light quark is also
non-relativistic.
\par

Let us give some details of the calculation of (\ref{6e}) and the corresponding
relations
(\ref{4e}). \underbar{Mutatis mutandis}, we consider the matrix element $<
0|V^{\mu}|0^+ >$ with the
$0^+$ wave functions given by (\ref{8e}). After some algebra, the matrix
element equivalent to
(\ref{15e}) is in the present case~:

\bea
&&< 0|V^{\mu}|0^+ >  = \sqrt{N_c}  \int {d{\bf p}_2 \over (2\pi)^3}
\sqrt{{\Sigma p^0_j \over M_0}}
\prod_{i=1,2} {\sqrt{k^0_i} \over \sqrt{p^0_i}}  \sqrt{{m_1 \over p^0_1}}
\sqrt{{m_2 \over
p^0_2}} \varphi^{(n)}_{1/2)}({\bf k}_2) \nn \\
 &&{1 \over \sqrt{6}} \hbox{\sixteenrm Tr} \left \{ \gamma^{\mu} \gamma_5 {\bf
B}_u
{\bf B}_{k_2} {1+ \gamma^0 \over 2} {\bf B}^{-1}_u {\bf B}_u (\boldsymbol{\sigma} \cdot
{\bf k}_2)
{\bf B}^{-1}_u {\bf B}_u {\rm {1 + \gamma^0 \over {2}}} {\bf B}^{-1}_{k_1} {\bf
B}^{-1}_u \right \}\
.	 \label{21e} \eea

\noi Using (\ref{16e}) and taking the limit (\ref{19e}), we obtain

\bea
&&< 0|V^{\mu}|0^+ >  =  {\sqrt{N_c} \over 4\sqrt{6}}{1 \over \sqrt{2v^0}} \int
{d{\bf p}_2
\over (2\pi)^3}  {1 \over p^0_2}{\sqrt{p_2 \cdot v} \over \sqrt{p_2 \cdot
v+m_2}}
\varphi^{(n)}_{1/2}({\bf k}_2) \nn \\
	&& \hbox{\sixteenrm Tr} \left [ - \gamma^{\mu} \left ( m_2+ {/ \hskip - 2
truemm p}_2 \right ) (1+{/ \hskip - 2 truemm} v) \left ( - {/ \hskip - 2 truemm
v} {/ \hskip - 2
truemm p}_2 +p_2 \cdot v \right )(1-{/ \hskip - 2 truemm v}) \right ]
\label{22e} \eea

\noi and using the fact that this expression must be proportional to $v^{\mu}$,
we obtain expression
(\ref{7e}), scaling invariant in the heavy quark limit. Moreover, varying the
current and considering
instead the matrix element $< 0|A^{\mu}|1^+_{1/2}>$ one can easily verify the
first relation
(\ref{4e}). \par

The result $g^{(3/2)} = 0$, that must hold in the heavy quark limit \cite{10r}
deserves a few
details. The matrix element to be considered is ($m$ denotes the polarization
of the state $1^+$, and
$n$ the radial excitation)~:

\bea
&&< 0|A^{\mu}|1^+_{3/2}, m > = \sqrt{N_c} \int {d{\bf p}_2 \over (2\pi)^3}
\sqrt{{\Sigma p^0_j \over M_0}}  \prod_{i=1,2} {\sqrt{k^0_i} \over
\sqrt{p^0_i}}
\sqrt{{m_1 \over p^0_1}} \sqrt{{m_2 \over p^0_2}} \nn \\
&&\hbox{\sixteenrm Tr} \left \{ \left ( {1+\gamma^0 \over 2} \sigma_2 {\bf
B}^{-1}_{p_1}
i\gamma^0 \gamma_5 \gamma^0 \gamma^{\mu} \gamma_5 {\bf B}_{p_2} {1+\gamma^0
\over 2} \right )
\left [ D({\bf R}_2) \right ] \left [ \varphi^{(n)}({\bf k}_2,m) \right ]^t
\left [ D^t({\bf R}_1)
\right] \right \} \nn \\
											\label{23e}
\eea

\noi where the wave function is, for the $1^+ j = \displaystyle{{3 \over 2}}$
states \cite{6r}~:

\beq
\varphi_{s_1,s_2}^{(n)}({\bf k}_2,m) = {i \over \sqrt{2}} {\bf e}^{(m)} \cdot
\left [ \sqrt{{2
\over 3}} {\bf k}_2 - {i \over \sqrt{6}} \left ({\bf k}_2 \times \boldsymbol{\sigma}
\right ) \right ] \sigma_2
\ \varphi^{(n)}_{3/2}({\bf k}_2)		 \label{24e}
\eeq

\noi where ${\bf e}^{(m)}$ is a unit vector, and the rotational invariant
function
$\varphi^{(n)}_{3/2}({\bf k}_2)$ is nor\-ma\-li\-zed according to (\ref{8e}).
The equivalent
expression to (\ref{21e}) will be

\bea
&&< 0|A^{\mu}|1^+_{3/2}, m > = \sqrt{N_c} \int {d{\bf p}_2 \over (2\pi)^3}
\sqrt{{\Sigma p^0_j \over M_0}} \prod_{i=1,2} {\sqrt{k^0_i} \over \sqrt{p^0_i}}
\sqrt{{m_1 \over p^0_1}} \sqrt{{m_2 \over p^0_2}} {\varphi^{(n)}_{3/2}({\bf
k}_2) \over \sqrt{2}} \nn
\\  &&\hbox{\sixteenrm Tr} \left \{ \gamma^{\mu} {\bf B}_u {\bf B}_{k_2} {1+
\gamma^0 \over 2}
{\bf e}^{(m)} \cdot \left [ \sqrt{{2 \over 3}} {\bf k}_2 + {i \over \sqrt{6}}
({\bf k}_2 \times
\boldsymbol{\sigma} ) \right ] {1+\gamma^0 \over 2} {\bf B}^{-1}_{k_1} {\bf B}^{-1}_u
\right \}		 \label{25e}
\eea

\noi We need to compute the expression ${\bf B}_u {\bf e}^{(m)} \cdot \left [
\sqrt{{2 \over
3}} {\bf k}_2 + {i \over \sqrt{6}} ({\bf k}_2 \times \boldsymbol{\sigma} ) \right ] {\bf
B}^{-1}_u$ in
the heavy quark limit. After some algebra, one gets~:

\beq
{\bf B}_u \  {\bf e}^{(m)} \cdot \left [ \sqrt{{2 \over 3}} {\bf k}_2 + {i
\over \sqrt{6}}
({\bf k}_2 \times \boldsymbol{\sigma} ) \right ] {\bf B}^{-1}_u \to {1 \over \sqrt{6}}
\left [ - {/ \hskip - 2
truemm \varepsilon}^{(m)}_v {/ \hskip - 2 truemm p}_2 + {/ \hskip - 2 truemm
\varepsilon}^{(m)}_v {/ \hskip - 2 truemm v} (p_2 \cdot v) -
\varepsilon^{(m)}_v \cdot p_2
\right ]    \label{26e} \eeq

\noi where $\varepsilon^{(m)}_v$ is the axial meson polarization in the heavy
quark limit. Using
(\ref{16e}) in the limit (\ref{19e}) we obtain finally,

\bea
&&< 0|A^{\mu}|1^+_{3/2}, m > =  {\sqrt{N_c} \over 8\sqrt{3}} {1 \over
\sqrt{2v^0}}
\int {d{\bf p}_2 \over (2\pi)^3}  {1 \over p^0_2} {\sqrt{p_2 \cdot v} \over
\sqrt{p_2 \cdot v+m_2}}
\varphi_{3/2}({\bf k}_2) \nn \\
&&\hbox{\sixteenrm Tr} \left \{ \gamma^{\mu} \left ( m_2+ {/ \hskip
- 2 truemm p}_2 \right ) (1+{/ \hskip - 2 truemm v}) \left [- {/ \hskip - 2
truemm
\varepsilon}^{(m)}_v {/ \hskip - 2 truemm p}_2 + {/ \hskip - 2 truemm
\varepsilon}^{(m)}_v {/
\hskip - 2 truemm v} (p_2 \cdot v) - \varepsilon^{(m)}_v \cdot p_2 \right
](1+{/ \hskip - 2 truemm
v}) \right \} 		 \label{27e}
 \eea

\noi This expression is covariant and satisfies heavy quark scaling. In the
rest frame,
particularizing to $m = 0$, one gets, because of rotational invariance of
$\varphi_{3/2}({\bf
k}_2)$~:

\bea
&&< 0|A^z|1^+_{3/2}, {\bf P} = {\bf 0}, m=0 > =  {\sqrt{N_c} \over \sqrt{3}} {1
\over\sqrt{2v^0}}
\int {d{\bf k}_2 \over (2\pi)^3}  {1 \over k^0_2} {\sqrt{k^0_2} \over
\sqrt{k^0_2 +m_2}}
\varphi_{3/2}({\bf k}_2) \nn \\
&&\left \{ - \left [ (k^x_2)^2 + {(k^y_2)}^2 \right ] + 2 {(k^z_2)}^2 \right \}
= 0		\ \ \ .
\label{28e}
\eea

Therefore the second relation (\ref{4e}), $g^{(3/2)} = 0$, follows. The
vanishing of $g^{(3/2)}$ can
be seen also from the following covariant argument. Contracting (\ref{27e})
with the four-vector
$\varepsilon^{(m)}_v$ we see that $g^{(3/2)}$ is proportional to the integral

\beq
g^{(3/2)} \sim \int {d{\bf p}_2 \over (2\pi)^3}  {1 \over p^0_2} {\sqrt{p_2
\cdot v}
\over \sqrt{p_2 \cdot v+m_2}} \varphi_{3/2}({\bf k}_2) \left [
3(\varepsilon^{(m)}_v \cdot p_2)^2 +
(v \cdot p_2)^2 - m_2^2 \right ]  \ .     	 \label{29e}
\eeq

\noi Two types of integrals appear :

\bea
&&I_{\mu \nu} =  \int {d{\bf p}_2 \over (2\pi)^3}  {1 \over p^0_2} {\sqrt{p_2
\cdot v}
\over \sqrt{p_2 \cdot v+m_2}} \varphi_{3/2}({\bf k}_2) \ p_{2\mu} \ p_{2\nu}
	 \label{30e} \\
&&I =  \int {d{\bf p}^2 \over (2\pi)^3}  {1 \over p^0_2} {\sqrt{p_2 \cdot v}
\over \sqrt{p_2
\cdot v+m_2}} \varphi_{3/2}({\bf k}_2) 	 \ \ \ . \label{31e}
\eea

\noi From covariance it follows that

\bea
	&&I_{\mu \nu} = A v_{\mu} v_{\nu} + B g_{\mu \nu} \nn \\
	&&I = C
\label{32e}
\eea

\noi where $A$, $B$, $C$ are constants. Contracting $I_{\mu \nu}$ with $g_{\mu
\nu}$ it follows

\beq
	4B+A = m_2^2 C
\label{33e}
\eeq

\noi and therefore

\beq
	g^{(3/2)} \sim  4B+A - m_2^2 C = 0 \ \ \ .
\label{34e}
\eeq

\section{Sum rules}
\hspace*{\parindent}

	Let us now show that the QCD heavy quark limit sum rules \cite{10r}

\beq
		X(w) \equiv  \sum_n {f^{(n)} \over f^{(0)}} \xi^{(n)}(w) = 1					\label{35e}
\eeq

\beq
		T_{1/2}(w) \equiv \sum_n  {f^{(n)}_{1/2} \over f^{(0)}} \tau^{(n)}_{1/2}(w) =
{1 \over 2}
\label{36e}
\eeq

\noi are satisfied within the present B-T scheme. \par

In order to prove Bjorken sum rule, in ref. \cite{5r} we used the completeness
relation at
\underbar{fixed} ${\bf P}$ that holds in the B-T formalism for the internal
wave functions. We
proceed here in the same way. We need to compute expressions of the form~:

\beq
	\sum_n < 0|\widetilde{O}|{\bf P}',n > < {\bf P}',n| O |{\bf P},0 >
\label{37e}
\eeq

\noi where $< 0|\widetilde{O}|{\bf P}',n >$ and $< {\bf P}',n| O |{\bf P},0 >$
will be related
respectively to the decay constants and to the Isgur-Wise functions. We need
${\bf P}'$ different
from ${\bf P}$ in order to demonstrate the sum rules (\ref{35e})-(\ref{36e})
for any value of the
scaling variable $w$.  \par

To obtain the sum rule (\ref{35e}) for $S$-waves, the simplest way is to take
${\bf P}' =
0$ for the intermediate states and choose the currents $O = \gamma^0
\gamma^{\nu}$, and
$\widetilde{O} = \gamma^0\gamma^{\mu}\gamma_5$ with $\mu = 0$.  Then, only the
$0^-$ intermediate
states contribute, because the $1^+$ states do not at ${\bf P}' = 0$, since
then the time
component of the polarization $\varepsilon^{(m)}_0 = 0$. Alternatively, to
obtain the sum rule
(\ref{36e}) for $P$-waves, we will take ${\bf P}' = 0$ and $O = \gamma^0
\gamma^{\nu}\gamma_5$, and
$\widetilde{O} = \gamma^0\gamma^{\mu}$ with $\mu = 0$.  Then, only the $0^+$
intermediate states
contribute, since the $1^-$ do not at ${\bf P}' = 0$.  \par

We start from the completeness
relation at fixed ${\bf P}'$~:

\beq
\sum_n \Psi^{(n)}_{s''_1s''_2} \left ( {\bf P}'-{\bf p}''_2,{\bf p}''_2 \right
)
\Psi^{(n)_*}_{s'_1s'_2} \left ( {\bf P}'- {\bf p}'_2,{\bf p}'_2 \right ) =
\delta_{s'_1s''_1} \
\delta_{s'_2s''_2} \ (2\pi)^3 \ \delta ({\bf p}'_2-{\bf p}''_2)	\ \ \ .
\label{38}
\eeq

\noi We need the matrix elements

\bea
&&< {\bf P}',n| O |{\bf P},0 > =  \int {d{\bf p}_2 \over (2\pi)^3}  \int {d{\bf
p}'_2 \over
(2\pi)^3}  \sum_{\{s_i\}} \sum_{\{s'_i\}} \nn \\
&&\Psi^{n({\bf P}')_*}_{s'_1s'_2}({\bf p}'_1,{\bf p}'_2) O_{s'_1s_1}({\bf
p}'_1,{\bf p}_1)
\Psi^{0({\bf P})}_{s_1s_2} ({\bf p}_1,{\bf p}_2) (2\pi)^3 \delta ({\bf
p}_2-{\bf p}'_2)
\delta_{s'_2s_2}
\label{39e}
\eea

\beq
< 0|\widetilde{O}|{\bf P},n > = \sqrt{N_c} \int {d{\bf p}_2 \over (2\pi)^3}
\sum_{\{s_i\}}
\widetilde{O}_{s_1s_2}({\bf p}_1,{\bf p}_2) \Psi^{n({\bf P})}_{s_2s_1} ({\bf
p}_1,{\bf p}_2)
\label{40e} \eeq

\noi related respectively to the IW functions and decay constants. In these
expressions~:

\beq
O({\bf p}'_1,{\bf p}_1) = {\sqrt{m_1m'_1} \over \sqrt{p^0_1p'^0_1}} {1+\gamma^0
\over 2}
{\bf B}^{-1}_{p'_1} \ O \ {\bf B}_{p_1} {1+\gamma^0 \over 2}
\label{41e}
\eeq

\beq
\widetilde{O}({\bf P}'-{\bf p}_2,{\bf p}_2) = \sqrt{{m_1 \over p'^0_1}}
\sqrt{{m_2 \over p^0_2}}
{1+\gamma^0 \over 2} \sigma_2 {\bf B}^{-1}_{p'_1} i\gamma^0\gamma_5
\widetilde{O} {\bf
B}_{p_2} {\rm {1+\gamma^0 \over {2}}}		 \label{42e}
\eeq

\noi and here $O$ and $\widetilde{O}$ are the Dirac matrices of the
corresponding currents.
After some algebra we obtain, using the completeness relation (38)~:

\bea
\label{43e}
&&\sum_n < 0|\widetilde{O}|{\bf P}',n > < {\bf P}',n| O |{\bf P},0 > =						\\
&&\sqrt{N_c}  \int {d{\bf p}_2 \over (2\pi)^3}  \sum_{\{s_i\}} \sum_{s'_1}
\widetilde{O}_{s'_1s_2}\left ( {\bf P}'-{\bf p}_2,{\bf p}_2 \right )
O_{s'_1s_1} \left (
{\bf P}'-{\bf p}_2,{\bf P}-{\bf p}_2 \right ) \Psi^{(0)}_{s_1s_2} \left ( {\bf
P}-{\bf p}_2,{\bf
p}_2 \right )  \nn \eea

\noi where the ground state wave function is given by \cite{1r}

\bea
&&\Psi^{(0)}_{s_1s_2}({\bf P}-{\bf p}_2,{\bf p}_2) = \sqrt{{\Sigma p^0_j \over
M_0}}  \prod_{i=1,2}
{\sqrt{k^0_i} \over \sqrt{p^0_i}} \sum_{\{s'_i\}} \left [ D({\bf R}_1) \right
]_{s_1s'_1}
\left [ D({\bf R}_2) \right ]_{s_2s'_2} \varphi^{(0)}_{s'_1s'_2}({\bf k}_2) \nn
\\
&&= \sqrt{{\Sigma p^0_j \over M_0}}  \prod_{i=1,2} {\sqrt{k^0_i} \over
\sqrt{p^0_i}}
\left [ D({\bf R}_1)\varphi^{(0)}({\bf k}_2) D^t({\bf R}_2) \right ]_{s_1s_2}
\label{44e}
\eea

\noi and using (\ref{6e}), in an obvious notation~:

\bea
&&\sum_n < 0|\widetilde{O}|{\bf P}',n > < {\bf P}',n| O |{\bf P},0 > = \nn \\
&&\sqrt{N_c}  \int {d{\bf p}_2 \over (2\pi)^3}  \sqrt{{\Sigma p^0_j \over M_0}}
 \prod_{i=1,2}
{\sqrt{k^0_i} \over \sqrt{p^0_i}}  {i \over \sqrt{2}} \varphi^{(0)}({\bf k}_2)
\nn \\
&&\hbox{\sixteenrm Tr} \left [ \widetilde{O}^t \left ( {\bf P}'-{\bf p}_2,{\bf
p}_2 \right )
O\left ( {\bf P}'-{\bf p}_2,{\bf P}-{\bf p}_2 \right ) D({\bf R}_1) \boldsymbol{\sigma}_2
D^t({\bf R}_2)
\right ] 				 \label{45e} \eea

\noi where the matrices ${\bf R}_i$ are given by (\ref{11e}). \par

To isolate the $0^-$ states, we take, as argued above~:

\bea
\label{46e}
&&\sum_n < 0|A^0|{\bf 0},n > < {\bf 0},n|V^{\nu}|{\bf P},0 > = \nn \\
&&\sqrt{N_c}  \int {d{\bf p}_2 \over (2\pi)^3}  \sqrt{{\Sigma p^0_j \over M_0}}
 \prod_{i=1,2}
{\sqrt{k^0_i} \over \sqrt{p^0_i}}   {\sqrt{m_1m'_1 \over p^0_1 p'^0_1}}
\sqrt{{m_1 \over p'^0_1}} \sqrt{{m_2 \over p^0_2}}  {i \over \sqrt{2}}
\varphi^{(0)}({\bf k}_2)
\\ &&\hbox{\sixteenrm Tr} \left \{ \left [ {1 + \gamma^0 \over 2} \sigma_2 {\bf
B}^{-1}_{p'_1} i\gamma^0 {\bf B}_{p_2} {1+\gamma^0 \over {2}} \right ]^t \left
[ {1+\gamma^0 \over 2}
{\bf B}^{-1}_{p'_1} \gamma^0 \gamma^{\nu} {\bf B}_{p_1} {1+\gamma^0 \over 2}
\right  ]
D({\bf R}_1)\sigma_2 D^t({\bf R}_2) \right \}			\nn						  \eea

Using $\sigma_2 ({\bf B}_p)^t \sigma_2  = {\bf B}^{-1}_p$, $\sigma_2
(\gamma^{\mu})^t \sigma_2  =
\gamma^{\mu}$, ${\bf B}_p \displaystyle{{1+\gamma^0 \over 2}} {\bf B}^{-1}_p =
\displaystyle{{1 \over
2}} \left ( 1 + \displaystyle{{{/ \hskip - 2 truemm p} \over m}} \right )$, the
fact that
$\displaystyle{{1 \over 2}}(1+\gamma^0)$ commutes with rotations, and the
relations

\beq
{\bf B}_v {\bf B}_{k_1}{\bf B}^{-1}_v = {m_1+ {/ \hskip - 2 truemm
p}_1 {/ \hskip - 2 truemm v} \over \sqrt{2m_1\left ( k_1^0+m_1 \right )}}
\qquad {\bf B}_v
{\bf B}^{-1}_{k_2} {\bf B}^{-1}_v = {m_2+{/ \hskip - 2 truemm v}{/ \hskip
- 2 truemm p}_2 \over \sqrt{2m_2 \left ( k_2^0 +m_2 \right )}}	 \label{47e}
\eeq

\noi one obtains, after some algebra, in the heavy quark limit $(p_1 \to  m_1v,
k_1^0 \to  m_1)$, for
the r.h.s. of (\ref{46e})~:

\bea
&&\sqrt{N_c} {1 \over \sqrt{v^0v'^0}} {1 \over \sqrt{v'^0}} \int {d{\bf p}_2
\over (2\pi)^3}
{1 \over p^0_2} \sqrt{m_2} \sqrt{k^0_2} {1 \over \sqrt{2}} \varphi^{(0)}({\bf
k}_2) \nn \\
&&\hbox{\sixteenrm Tr} \left \{ {1 \over 2} \left ( 1 + \gamma^0 \right )
\gamma^{\nu} {1 \over 2}
(1 + {/ \hskip - 2 truemm v}) {\bf B}_v {\bf B}_{k_1} {\bf B}^{-1}_{k_2} {\bf
B}^{-1}_v \right \} =
\nn \\
&&\sqrt{N_c} {1 \over \sqrt{4v^0v'^0}} {1 \over \sqrt{2v'^0}}  \int {d{\bf p}_2
\over (2\pi)^3}
{1 \over p^0_2}{\sqrt{p_2 \cdot v} \over \sqrt{p_2 \cdot v+m_2}} {1 \over
\sqrt{2}}
\varphi^{(0)}({\bf k}_2) {1 \over 2} 	 \\
&&\hbox{\sixteenrm Tr} \left \{ \left ( 1 + \gamma^0 \right ) \gamma^{\nu} (1 +
{/ \hskip - 2
truemm v}) \left ( m_2+{/ \hskip - 2 truemm p}_2 \right ) \right \} = {1 \over
\sqrt{4v^0v'^0}}
{1 \over \sqrt{2v'^0}}  \sqrt{M} f^{(0)} \left ( g^{0\nu} + v^{\nu} \right ) .
\nn \label{48e}
\eea

\noi On the other hand, the l.h.s. of (\ref{46e}) reads~:

\bea
&&\sum_n < 0|A^0|{\bf 0},n > < {\bf 0},n|V^{\nu}|{\bf P},0 > = \nn \\
&&{1 \over \sqrt{4v^0v'^0}} {1 \over \sqrt{2v'^0}} \sum_n \sqrt{M}  f^{(n)}_P
\xi^{(n)}(w)
\left ( g^{0\nu} + v^{\nu} \right )
\label{49e}
\eea

\noi and therefore the sum rule (\ref{35e}) follows. \par

The sum rule for the $P$-states (\ref{36e}) follows straightforwardly in a
similar manner by taking
${\bf P}' = 0$ and $O = \gamma^0\gamma^{\nu}\gamma_5$, $\widetilde{O} =
\gamma^0\gamma^{\mu}$ with
$\mu = 0$, since then only the $0^+$ intermediate states contribute, as pointed
out above. However,
to illustrate the methods of calculation in the BT formalism, we will now
verify the further sum rule
(\ref{36e}) from the general expressions for $f^{(n)}_{1/2}$ and
$\tau^{(n)}_{1/2}(w)$, without
appealing to any particular frame.  \par

We want to evaluate the expression

\beq
		T_{1/2}(w) f^{(0)} = \sum_n f^{(n)}_{1/2} \ \tau^{(n)}_{1/2}(w) \ \ \ .
\label{50e}
\eeq

\noi From the explicit expressions (\ref{7e}) for $f^{(n)}_{1/2}$ and
$\tau^{(n)}_{1/2}(w)$ from ref.
\cite{6r}~:
\bea
&&\tau^{(n)}_{1/2}(w) = {1 \over 2\sqrt{3}} \int {d{\bf p}'_2 \over (2\pi)^3}
{1 \over p'^0_2}
{\sqrt{(p'_2 \cdot v')(p'_2 \cdot v)} \over \sqrt{(p'_2 \cdot v+m_2)(p'_2 \cdot
v'+m_2)}}
\varphi^{(n)}_{1/2}({\bf k}'_2)^* \varphi^{(0)}({\bf k}''_2) \nn \\
&&\times {(p'_2 \cdot v')(p'_2.v+m_2)-(p'_2 \cdot v)(p'_2 \cdot
v+wm_2)+(1-w)m_2^2 \over 1-w}
\label{51e}
\eea

\noi with

\bea
&k'_2 = {\bf B}^{-1}_v p'_2  		&\qquad \qquad k''_2 = {\bf B}^{-1}_{v'} p'_2 \
\ \ .
\label{52e}
\eea

\noi The computation of the sum (\ref{50e}) leads to the expression

\beq
\sum_n \varphi^{(n)}_{1/2}({\bf k}'_2)^* \varphi^{(n)}_{1/2}({\bf p}_2) =
6\pi^2 {1 \over
{\bf p}^2_2 \ {\bf k}'^2_2}  \delta (|{\bf p}_2| - |{\bf k}'_2|)
\label{53e}
\eeq

\noi where the r.h.s. follows from ref. \cite{6r}. One can then perform the
integration over ${\bf
p}_2$ that amounts to replace all $p_2^0$ by $(p'_2 \cdot v)$. Realizing then
that

\bea
&& {(p'_2 \cdot v')(p'_2 \cdot v+m_2)-(p'_2 \cdot v)(p'_2 \cdot
v+wm_2)+(1-w)m_2^2 \over 1-w} = \nn
\\ &&{\bf k}'^2_2 \left ( -1 + {p'_2 \cdot v'-w(p'_2\cdot v) \over (p'_2 \cdot
v-m_2)(1-w)}
\right )						 \label{54e}
\eea

\noi one gets

\bea
&&2T_{1/2}(w) f^{(0)} = 2 \sum_n f^{(n)}_{1/2} \tau^{(n)}_{1/2}(w) = \nn \\
&&f^{(0)} - \sqrt{{2N_c \over \pi}}  \int {d{\bf p}'_2 \over (2\pi)3} {1 \over
p'^0_2}
{\sqrt{(p'_2 \cdot v')} \over \sqrt{(p'_2 \cdot v'+m_2)}} \varphi^{(0)}({\bf
k}''_2)
{p'_2 \cdot v-w(p'_2 \cdot v') \over 1-w}	 \ \ \ .  \label{55e}
\eea

\noi The second term involves the integral

\beq
 \int {d{\bf p}'_2 \over (2\pi)^3} {1 \over p'^0_2}{\sqrt{(p'_2 \cdot v')}
\over
\sqrt{(p'_2 \cdot v'+m_2)}} \varphi^{(0)}({\bf k}''_2) p'_{\mu} = C v'_{\mu}
\label{56e}
\eeq

\noi where $C$ is some constant, because of covariance. Inserting the r.h.s. of
(\ref{56e}) into
(\ref{55e}) we see that the second term of the r.h.s. of (\ref{55e}) vanishes
for $w \not= 1$, and
then the sum rule (\ref{36e}) follows. For $w = 1$, one can choose $v = (v^0,
{\bf v})$, $v' =
(1, {\bf 0})$ and make an expansion of $(w - 1)^{-1}$ for ${\bf v} \to {\bf
0}$. Then, the second
term in the r.h.s. of (\ref{55e}) vanishes from rotational invariance. The same
method allows also to
obtain the sum rule (\ref{35e}) along similar lines.

\section{Numerical results}
\hspace*{\parindent}

As explained at length in ref. \cite{1r}, the dynamics in the B-T formalism
depends on the form of
the mass operator $M$ at ${\bf P} = 0$. This mass operator can be of any form

\beq
			M = K(\{{\bf k}_i\}) + V(\{{\bf r}_i,{\bf p}_i\})
\label{57e}
\eeq

\noi and one obtains covariance of the form factors and IW scaling with only
the very general
assumption of rotational invariance of $M$. We are going now to give the
results for the decay
constants for various Ans\"atze of the operator $M$, not only va\-rious forms for
the potential
$V(\{{\bf r}_i,{\bf p}_i\})$, but also for the kinetic energy $K(\{{\bf
k}_i\})$, that can be taken
to be of the non-relativistic $\displaystyle{{{\bf k}^2_i \over 2m_i}}$  or
relativistic
$\sqrt{{\bf k}^2_i+m^2_i}$ forms. We choose such models in order to emphasize
the physics involved in
the decay constants, \underbar{sensitive to the short distance part of the
potential and to the
scaling behaviour of}\break \noindent \underbar{the kinetic energy} (quadratic
or linear in $k$). Of
course, any scheme that we adopt should give a reasonable fit of the whole
meson spectrum. The
success of quark models in the
description of the spectrum with either non-relativistic or relativistic
kinetic energies shows that the spectrum by itself does not constrain the form
of this
kinetic energy. Interestingly, we have shown in ref. \cite{7r} that quark
models of form factors in
the BT formalism show a clear preference for the relativistic form of the
kinetic energy, since they
give a slope of the elastic IW function $\rho^2 \cong 1$, while models with a
non-relativistic
kinetic energy give a larger, phenomenologically unacceptable value. \par

We list here a number of phenomenological quark models of the hadron spectrum
that we will use,
specifying their interesting features : \par \vskip 3 truemm

1) Isgur, Scora, Grinstein and Wise (ISGW) \underbar{spectroscopic model}
\cite{11r} (to be
distinguished from the ISGW non-relativistic \underbar{model of form
factors})~:
non-relativistic kinetic energy $\displaystyle{{{\bf k}^2_i  \over 2m_i}}$
with linear plus Coulomb
potential. \par \vskip 3 truemm

2) Veseli and Dunietz (VD) model \cite{12r} : relativistic kinetic energy
$\sqrt{{\bf k}^2_i+m^2_i}$ with linear plus Coulomb potential. The Coulomb part
is not
regularized. \par \vskip 3 truemm

3) Godfrey and Isgur model (GI) \cite{13r} : relativistic kinetic energy
$\sqrt{{\bf k}^2_i+m^2_i}$ with linear plus a regularized short distance part.
This scheme also
incorporates the fine structure of the potential. \par \vskip 3 truemm

4) Richardson potential with relativistic kinetic energy, as used by Colangelo,
Nardulli and Pietroni
(CNP) \cite{14r}. This model exhibits the asymptotic freedom behaviour at short
distances, but the
Coulomb singularity  (logarithmically corrected) is regularized by a cut at
small $r$. \par \vskip
3 truemm

	These models are solved numerically using a harmonic oscillator basis. In
Table 1 we give the
masses and decay constants of the states $n = 0$, $\ell = 0$ or 1  as a
function of $N_{max}$ in the
different models. $N_{max}$ means the number of radial excitations included in
the truncated basis,
the ground state gaussian plus up to $N_{max}$ radially excitated harmonic
oscillator wave
functions~: $N_{max}+1$ is then the dimension of the truncated Hilbert space.
In Table 2 we give the
decay constants of a number of radial excitations in the GI model.\par

Let us begin our discussion with the models with relativistic kinetic energy. A
first important
remark to make is that dimensional analysis implies that a Hamiltonian (or mass
operator in the B-T
formalism) with kinetic energy of the relativistic form $\sqrt{{\bf
k}^2_i+m^2_i}$ and a
Coulomb potential implies divergent wave functions at the origin
$\Psi^{(n)}(0)$ (or decay constants
$f^{(n)}$) for $S$-waves. The reason for this behaviour is that the kinetic and
the potential
energies exhibit the same scaling properties at large momentum or small
distances. The
relativistic kinetic energy is not efficient enough in smoothing the $r$-space
wave function. This
divergence of the wave function at the origin is the cause of the existence of
a critical coupling
$\alpha^{crit}$ in this class of models~: for $\alpha > \alpha^{crit}$ arises
the so-called
phenomenon of fall in the center. For a discussion, see for
example the paper by Hardekopf and Sucher \cite{15r}. On the contrary, the
decay constants for
$P$-wave mesons remain finite even in this case. A short distance Coulomb part
corrected by
asymptotic freedom exhibits the same phenomenon, although $f^{(n)}$ diverges
only logarithmically in
this latter case, instead of as a power in the former. These general features
are exhibited by the
model of Veseli and Dunietz \cite{12r}, as shown in the Table, and have been
underlined by these
authors. In our Table, the finite values obtained for $\sqrt{M}$ $f^{(0)}$
using the singular model
for $N_{max} = 10$, 15 or 20 are just an artifact of the truncation method.
\par

The GI model \cite{13r}, a model of the meson spectrum for all $q\bar{q}$,
$Q\bar{q}$ and $Q\bar{Q}$
systems, chooses a short distance part with a fully regularized Coulomb
singularity. Within this
model it is possible to compute both types of decay constants $f^{(n)}$ and
$f^{(n)}_{1/2}$. In the
Table we give the scaling invariant quantities $\sqrt{M} f^{(0)}$ and $\sqrt{M}
f^{(0)}_{1/2}$ for
$S$ and $P$-wave $n = 0$ mesons, computed in the heavy quark limit. \par

We have also made the calculation for the CNP model (Richardson potential plus
relativistic kinetic
energy) \cite{14r}. \par

In the case of a kinetic energy of the non-relativistic form
$\displaystyle{{{\bf k}^2_i \over
2m_i}}$ the $S$-wave decay constants are finite even in the presence of a
Coulomb singularity. This
feature is examplified by the ISGW \cite{11r} model in Table 1. Moreover, the
decay constants
are smaller in this case than for the models of relativistic kinetic energy,
another
manifestation of the singular behavior of the latter. \par

It is interesting to notice that in the heavy quark limit and for $N_{max} =
20$ one gets, for the
models with relativistic kinetic energy and regularized Coulomb singularity~:

\bea
&&\sqrt{M} f^{(0)} = 0.67 \ {\rm GeV}^{3/2} \ \hbox{(GI model)} \ ; \  0.83 \
{\rm GeV}^{3/2} \
\hbox{(CNP model)} \nn \\
&&\sqrt{M} f^{(0)}_{1/2} = 0.64 \ {\rm GeV}^{3/2} \ \hbox{(GI model)} \ ; \
0.70 \ {\rm
GeV}^{3/2} \ \hbox{(CNP model)} \ \ \ .		 \label{58e}
\eea

\noi Applying this \underbar{asymptotic} result to the $B$ meson, one obtains

\beq
			f_B \cong  \ \hbox{300 MeV - 350 MeV}	\ \ \ .
\label{59e}
\eeq

\noi This value is not far away, although slightly larger than the values
obtained in
\underbar{lattice QCD in the static limit}, to which it should naturally be
compared, which ranges
between 220 and 290 MeV \cite{16r}. However, one should keep in mind that the
lattice QCD result
includes logarithmic corrections absent in our phenomenological scheme. \par

It is worth noticing that we obtain the same order of magnitude for the decay
constant
$f^{(0)}_{1/2}$ . This is phenomenologically important because, reasoning
within the factorization
assumption, this means that the \underbar{emission} of $D^{**}(0^+)$ and
$D^{**}(1^+, j =
\displaystyle{{1 \over 2}})$ is expected to be important in $B$ decays. On the
contrary, the
emission of $D^{**}(j = \displaystyle{{3 \over 2}})$ will be suppressed. Table
2 shows that the
decay constants of radial excitations are of the same order of magnitude as in
the ground state.
However, the error due to the truncation is larger as $n$ increases. \par

Finally, let us study the convergence of the sum rules (\ref{35e}) and
(\ref{36e}), that we have
shown formally to hold, in models that give finite results (GI, CNP and ISGW)
for which we can
compute the decay constants $f^{(n)}$, $f^{(n)}_{1/2}$ and the IW functions
$\xi^{(n)}(w)$,
$\tau^{(n)}_{1/2}(w)$ \cite{8r}. The convergence of the Bjorken-Isgur-Wise sum
rule \cite{4r} has
been studied in ref. \cite{8r}. \par

Let us define

\beq
X^{(N)}(w) = \sum_{n=0}^N {f^{(n)} \over f^{(0)}} \xi^{(n)}(w) 	\qquad \qquad
T^{(N)}_{1/2}(w) =
\sum_{n=0}^N {f^{(n)}_{1/2} \over f^{(0)}} \tau^{(n)}_{1/2}(w) \ \ \ .
\label{60e}
\eeq

\noi We compute the sums for $N = N_{max}$ in the different models for various
values of $w$, and see
how they compare to the r.h.s. of the sum rules (\ref{35e}) and (\ref{36e}).
Let us recall that
$N_{max}$ is the maximal number of radial excitations included in the
truncation method (the
dimension of the variational base is $N_{max}+1$). We show the results for the
Godfrey-Isgur model in
Figures 1 and 2. The $Ox$ axis represents $1/(N_{max}+1$). We observe that
these sums converge fairly
well towards the r.h.s. of the sum rules (respectively 1 and $\displaystyle{{1
\over 2}}$) as we
increase $N_{max}$. For fixed $N_{max}$ we can ask how the partial sums
$X^{(N)}(w)$ and
$T^{(N)}_{1/2}(w)$ ($N \leq N_{max}$) behave as a function of $N$, i.e. how
fast $X^{(N)}(w)$,
$T^{(N)}_{1/2}(w)$  approach $X^{(N_{max})}(w)$, $T^{(N_{max})}_{1/2}(w)$ when
$N$ increases ($N \geq
0$). Let us give the example $N_{max} = 20$. The convergence is rather fast,
but it degrades as $w$
increases. Concerning the $X$ sum rule, for $w = 1$ one has trivially
$X^{(0)}(1) =
X^{(N_{max})}(1)$, because the ground state saturates the sum, since
$\xi^{(n)}(1) = 0$ ($n > 0$).
As one increases $w$, one one needs to sum up to $N = 3$, 4, 5, 6 or 7
respectively for $w = 1.1$,
1.2, 1.3, 1.4 or 1.5 to approach $X^{(N_{max})}(w)$ at the 5~$\%$ level.
Concerning the $T_{1/2}$ sum
rule, one needs $N = 3$, 4, 5, 6, 7 or 8 respectively for $w = 1.1$, 1.2, 1.3,
1.4 or 1.5 to
approach $T^{(N_{max})}_{1/2}(w)$ at the 10~$\%$ level. \par

For the CNP model and the non-relativistic ISGW model the convergences both in
$N_{max}$ and
in $N$ for fixed $N_{max}$ are not as good. For the case of the VD model one
gets finite
$f^{(n)}_{1/2}$ and $\tau^{(n)}_{1/2}(w)$, but the convergence toward the
r.h.s. of the sum rule
(\ref{30e}) does not improve as one increases $N_{max}$. This results from the
divergence of the
denominator $f^{(0)}$ in (\ref{35e}), (\ref{36e}) when $N_{max} \to \infty$.
\par

In conclusion, we have studied the decay constants of heavy-light mesons in the
heavy mass limit in
a class of models \`a la Bakamjian and Thomas, with different Ans\"atze for the
dynamics at rest, with
non-relativistic or relativistic kinetic energies. Each particular model gives
an acceptable
phenomenological description of the spectrum. The models with relativistic
kinetic energy, that
yield a slope of the elastic IW function $\rho^2 \cong 1$ (as shown in ref.
\cite{7r}), give finite
decay constants if the Coulomb singularity of the potential is regularized, as
in the GI model. At
the $B$ mass, one finds $f_B$ slightly larger than in the static limit of
lattice QCD. The decay
constants of $D^{**}$ with $j = \displaystyle{{1 \over 2}}$  are of the same
order of magnitude.
Moreover, we have shown that heavy quark limit sum rules involving decay
constants \cite{10r} are
satisfied by these class of quark models \`a la Bakamjian and Thomas, and in the
case of the
Godfrey-Isgur model the convergence of the sum rules is quite fast.

\newpage
\section*{Table captions}
\noi {\bf \underbar{Table 1.}} Decay constants of $n = 0$, $\ell = 0$ and  $n =
0$, $\ell = 1$, $j =
\displaystyle{{1 \over 2}}$  mesons in the various models. The ISGW is
non-relativistic. In the VD,
GI and CNP models, the kinetic energy is relativistic. In the VD model the
$S$-wave decay constants
diverge due to the Coulomb singularity. In the GI and CNP models the Coulomb
singularity is
regularized and the decay constants are finite. $N_{max}$ stands for the number
of radial
excitations included in the truncated  variational harmonic oscillator basis.
\par \vskip 5 truemm

\noi {\bf \underbar{Table 2.}} Decay constants $f^{(n)}$ and $f^{(n)}_{1/2}$
for the first radial
excitations of $\ell = 0$ and  $\ell = 1$, $j = \displaystyle{{1 \over 2}}$
mesons. The error in
parenthesis is estimated by comparing the number of radial excitations included
in the truncated
variational harmonic oscillator basis $N_{max}$ = 20 and $N_{max}$ = 10. For $n
\ \gsim \ 5$ the
error becomes larger than 20 $\%$. \par \vskip 1 truecm

\section*{Figure captions}

\noi {\bf \underbar{Figure 1.}} Convergence of the heavy quark limit $S$-wave
sum rule (\ref{35e}) in
the GI model \cite{13r} for different values of the scaling variable $w$ as one
increases $N_{max}$,
the number of radial levels included in the truncated variational harmonic
oscillator basis. The $Ox$
axis represents $(N_{max}+1)^{-1}$ and the $Oy$-axis
$\sum\limits_{n=0}^{N_{max}}
\displaystyle{{f^{(n)} \over f^{(0)}}} \xi^{(n)}(w)$, that the sum rule
predicts to be equal to 1 for
$N_{max} \to \infty$. The different lines correspond to $w = 1.0$, 1.1, 1.2,
1.3, 1.4, 1.5, from up
to down. \par \vskip 5 truemm

\noi {\bf \underbar{Figure 2.}} Convergence of the heavy quark limit $S$-wave
sum rule (\ref{36e}) in
the GI model \cite{13r} for different values of the scaling variable $w$ as one
increases $N_{max}$,
the number of radial levels included in the truncated variational harmonic
oscillator basis. The
$Ox$ axis represents $(N_{max}+1)^{-1}$ and the $Oy$-axis
$\sum\limits_{n=0}^{N_{max}}
\displaystyle{{f^{(n)}_{1/2} \over f^{(0)}}} \tau^{(n)}_{1/2}(w)$, that the sum
rule predicts to be
equal to $\displaystyle{{1 \over 2}}$ for $N_{max} \to \infty$. The different
lines correspond to
$w = 1.0$, 1.1, 1.2, 1.3, 1.4, 1.5, from up to down.

\newpage
\begin{center}
\begin{tabular}{|c|c|c|c|c|c|c|}
\hline
MODEL	&$M_Q$ &$N_{max}$ &$M^{(0)}$-$M_Q$ &$M^{(0)}_{1/2}$-$M_Q$ &$\sqrt{M}
f^{(0)}$ &$\sqrt{M}
f^{(0)}_{1/2}$ \\
&(GeV)	& &(GeV)	 &(GeV)	&(GeV$^{3/2}$)	 &(GeV$^{3/2}$) \\
\hline
ISGW [11]	&10$^{4}$	&10	&0.0438	&0.5467	&0.422	&0.235 \\
	&10$^{4}$	&15	&0.0436	&0.5467	&0.428	&0.235 \\
	&10$^{4}$	&20	&0.0435	&0.5467	&0.431	&0.236 \\
	&	&infinite	&	&	&finite	&finite \\
\hline
VD [12]	&10$^{4}$	&10	&0.119	&0.620	&1.36	&0.603 \\
	&10$^4$	&15	&0.108	&0.620	&1.58	&0.617 \\
	&10$^{4}$	&20	&0.108	&0.620	&1.76	&0.631 \\
 &		&infinite	&	&	&infinite	&finite \\
\hline
GI [13]	&10$^{4}$	&10	&0.386	&0.792	&0.649	&0.620 \\
	&10$^{4}$	&15	&0.386	&0.792	&0.662	&0.632 \\
	&10$^{4}$	&20	&0.386	&0.792	&0.667	&0.640 \\
	&10$^{4}$	&infinite		&	& &finite	&finite \\
\hline
CNP [14]	&10$^{4}$	&10	&0.389	&0.859	&0.747	&0.669 \\
	&10$^{4}$	&15	&0.387	&0.858	&0.798	&0.691 \\
	&10$^{4}$	&20	&0.386	&0.858	&0.828	&0.704 \\
	&	&infinite	&	&	&finite	&finite \\
\hline
\end{tabular} \par
\vskip 5 truemm
{\bf Table 1}
\end{center}

\vskip 1 truecm
\begin{center}
\begin{tabular}{|c|c|c|}
\hline
radial excitation	&$\sqrt{M} f^{(n)}$ &$\sqrt{M} f^{(n)}_{1/2}$ \\
&(GeV$^{3/2}$)	 &(GeV$^{3/2}$) \\
\hline
n = 0	    &0.67(2)	    &0.64(2)	\\
n = 1	    &0.73(4)	    &0.66(4)	\\
n = 2	    &0.76(5)	    &0.71(5)	\\
n = 3	    &0.78(9)	    &0.73(8) \\
n = 4	    &0.80(10)	    &0.76(11) \\
n = 5	    &0.81(17)	    &0.77(17) \\
n = 6	    &0.82(15)	    &0.78(15)	\\
n = 7	    &0.82(28)	    &0.78(27)	\\
n = 8	    &0.83(25)	    &0.79(25)	\\
n = 9	    &0.80(40)	    &0.76(40) \\
n = 10	    &0.83(42)	    &0.79(40) \\
\hline
\end{tabular} \par
\vskip 5 truemm
	{\bf Table 2}
\end{center}

\newpage
\begin{figure}[p]
\vspace{-60mm}
$$\hspace{-10mm}
\epsfysize=200mm\epsfbox{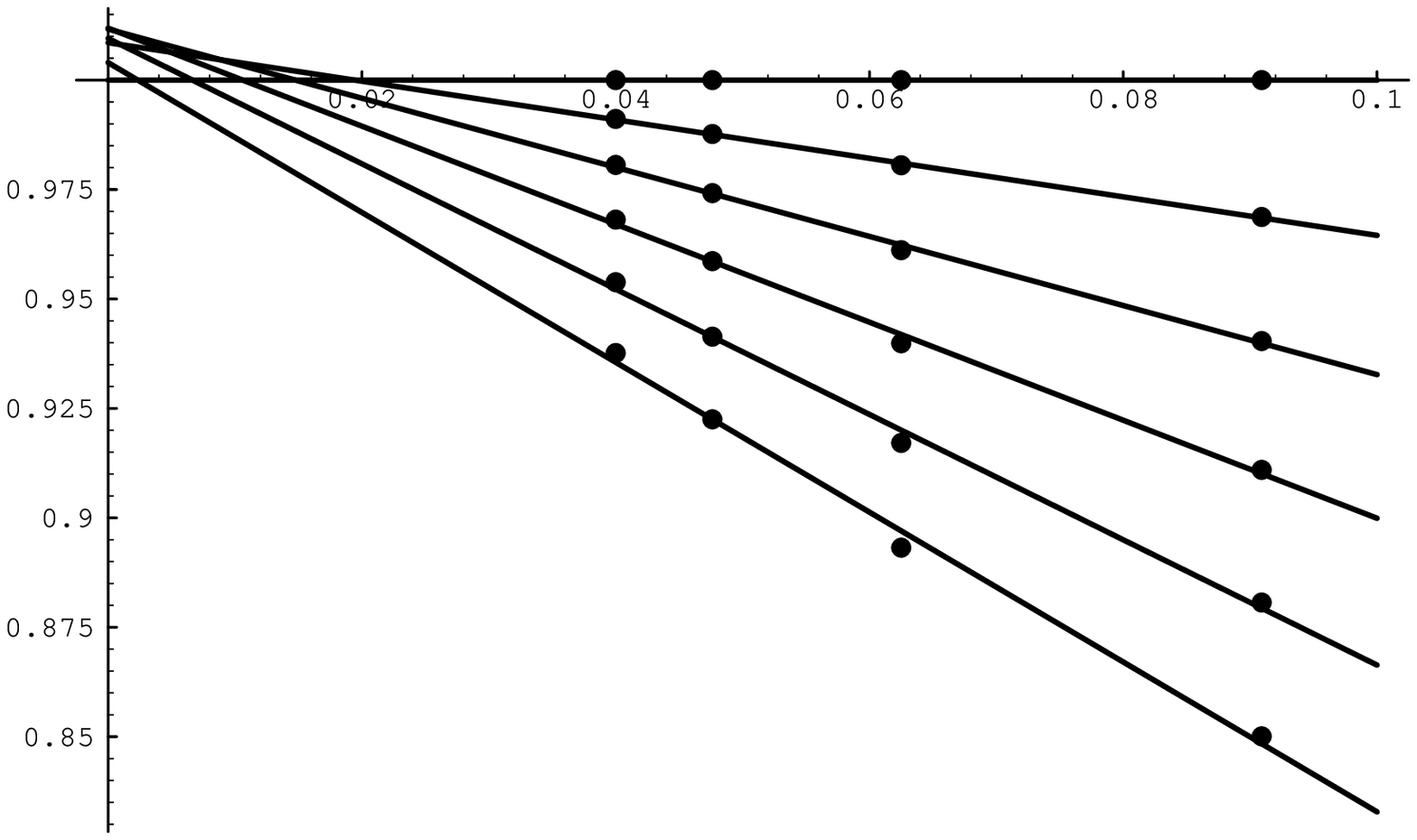}$$\vspace{-60mm}
\caption{}
\end{figure}
\begin{figure}[p]
\vspace{-60mm}
$$\hspace{-10mm}
\epsfysize=200mm\epsfbox{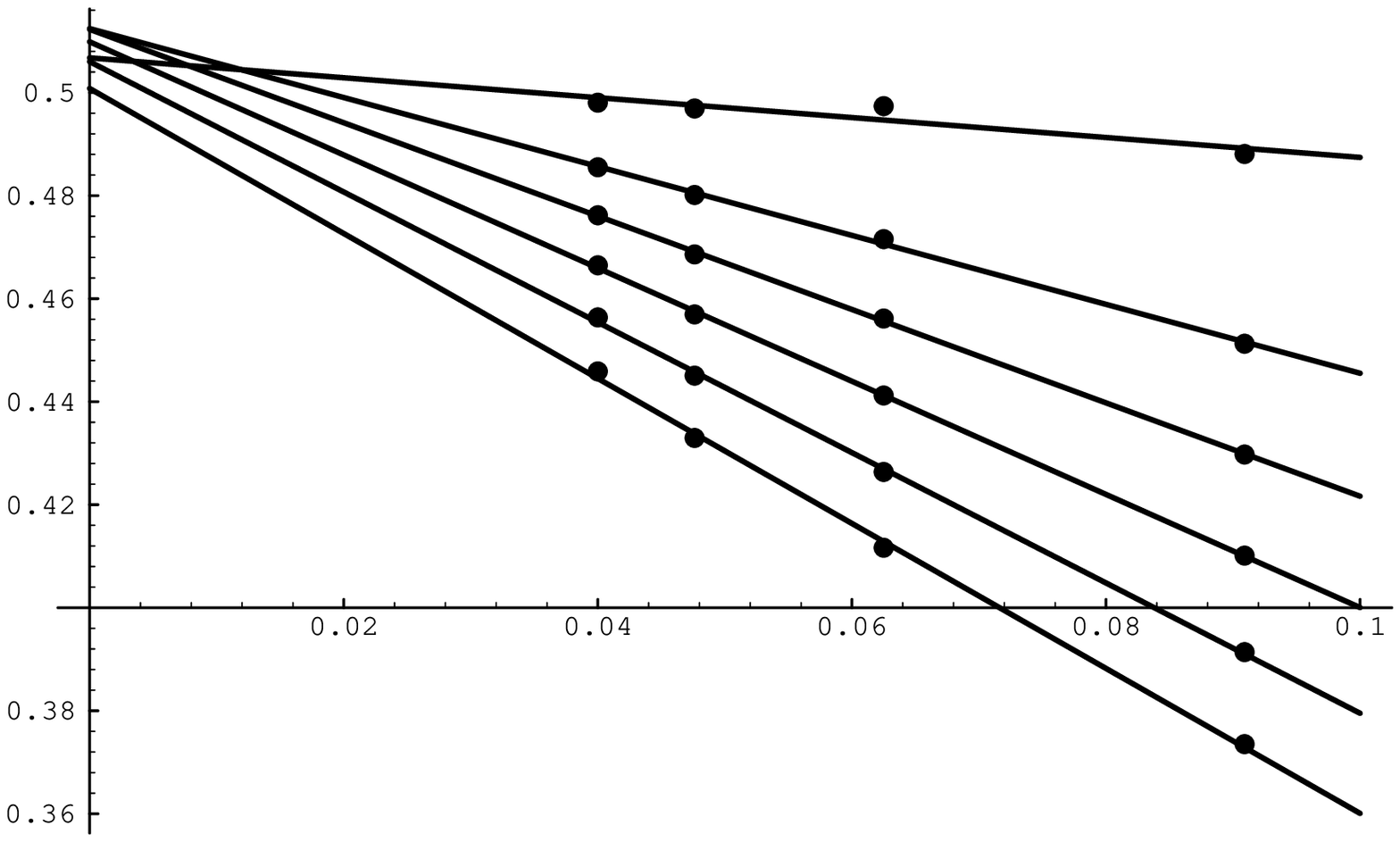}$$\vspace{-60mm}
\caption{}
\end{figure}

\end{document}